# Single-Shot Two Dimensional Time Resolved Coherent Anti Stokes Raman Scattering

**Yuri Paskover, I.Sh. Averbukh and Yehiam Prior**
*Department of Chemical Physics, Weizmann Institute of Science, Rehovot, Israel 76100*
*yehiam.prior@weizmann.ac.il*

**Abstract:** Single-shot time resolved Coherent Anti-Stokes Raman Scattering (CARS) is presented as a viable method for fast measurements of molecular spectra. The method is based on the short spatial extension of femtosecond pulses and maps time delays between pulses onto the region of intersection between broad beams. The image of the emitted CARS signal contains full temporal information on the field-free molecular dynamics, from which spectral information is extracted. The method is demonstrated on liquid samples of $CHBr_3$ and $CHCl_3$ and the Raman spectrum of the low-lying vibrational states of these molecules is measured.

Femtosecond nonlinear optical spectroscopy has emerged as a powerful tool for tracing intramolecular processes as they occur. The ability to excite and probe molecules on time scales faster than their rotational and vibrational motion opened up new directions in the analysis of structure as well as dynamics of intramolecular processes. In analogy to NMR, optical two- dimensional (2D) spectroscopic methods have been introduced which are geared towards the retrieval of couplings between different degrees of freedom of molecules [1]. In these experiments, ultrashort pulses were used to study vibrational as well as electronic degrees of freedom either by direct (IR) or Raman type excitation and probing [2-10]. Coupling between degrees of freedom is expressed as correlations between different excitations of molecular states. Since the molecular response has to be probed before dephasing occurs, most experiments are performed in the time domain. Thus, one usually changes the timing (or in some cases the spectral contents) of the exciting pulses, and the probing is usually done by scanning the delay between the excitation and the probe pulse. The two time delays provide access to the two relevant spectroscopic dimensions.

In Time Delayed CARS (TDCARS), the first two pulses (pump and Stokes) establish coherence between vibrational states of a molecule, and a third, delayed, probe pulse is scattered off this coherence. In resonant CARS experiments, an additional delay between the pump and Stokes pulses enables evolution of the prepared wavepacket on the excited electronic state potential surface [11], allowing selectivity in the preparation of ground state molecular vibrational states and providing, in fact, a two-dimensional picture of correlation between ground and excite electronic states' vibrational wavepackets [12].

Most time domain 2D experiments require scanning of at least two delays, leading to long measurement times, and imposing severe (sometime prohibitive) requirements on laser system and molecular long term stability. In the present communication we demonstrate an experimental technique that enables single-shot measurements of time resolved CARS by a one-to-one mapping of delay times to spatial coordinates in the intersection region of the three input beams. In this technique, various delay time combinations are realized all together, and as demonstrated, the entire measurement is performed in a single shot, enabling the tracking of intramolecular dynamics and coupling between different degrees of freedom.

Single-shot CARS was implemented many years ago, with long (nanosecond) pulses, for spatial and spectroscopic characterization of gas jets [13] or of laser produced plasmas [14, 15]. With femtosecond pulses the "spatial extent" of a pulse may be much shorter than the lateral dimensions of the laser beam, opening possibilities for using the arrival time of the pulses to a specific point in space for temporal resolution of the nonlinear optical signal.

Encoding the interpulse delay on the spatial axes of the beams' intersection area have been pioneered (in picosecond pump-probe experiments) by Malley and Rentzepis [16]. Various beam intersection geometries were used for pulse diagnostics [17], and analyzed extensively by Fourkas et al. [18]  Other single-shot arrangements were based on pump-probe measurements with white light continuum [19].  Zamith et al. [20] used the instantaneous frequency of a chirped pulse to map frequency to time delays in pump probe experiments. Recently Pouline and Nelson [21] have demonstrated encoding of interpulse delay on different propagation directions of a focused beam. DeCamp and Tokmakoff [22] utilized the broad bandwidth of an ultrashort pulse to map spectral (rather than temporal) resolution on the spatial dimensions of a sample.

In this work, we demonstrate experimentally that coherent field-free evolution of vibrational wavepackets can be captured and retrieved by the spatial imaging of a CARS signal produced by laser pulses that are spatially shorter than the lateral dimensions of their beams. Consider a Degenerate Four Wave Mixing experiment in a three-dimensional Folded Boxcars configuration[23].  Three input beams pass through three corners of a square, and the signal beam is emitted toward the fourth corner due to phase matching conditions (Figure 1).

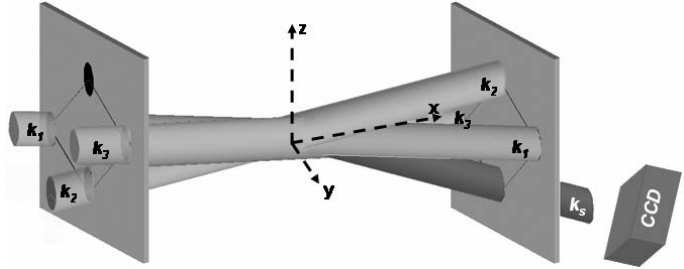

**Figure 1 Configuration of the three incoming beams in single-shot CARS experiment. The signal beam is collected directly on a camera.**

The geometrical axes are defined as follows: the x-axis lies along the bisector between beams 1 and 3, the y-axis is in the same plane and perpendicular to it, the z-axis points upwards. The directions of propagation of all three beams along the $k_1$, $k_2$ and $k_3$ axes are given by:

$$\vec{k}_1 = \begin{pmatrix} \cos\theta \\ \sin\theta \\ 0 \end{pmatrix}; \; \vec{k}_2 = \begin{pmatrix} \cos\theta \\ 0 \\ \sin\theta \end{pmatrix}; \; \vec{k}_3 = \begin{pmatrix} \cos\theta \\ -\sin\theta \\ 0 \end{pmatrix}; \; \vec{k}_s = \vec{k}_1 - \vec{k}_2 + \vec{k}_3 = \begin{pmatrix} \cos\theta \\ 0 \\ -\sin\theta \end{pmatrix}. \qquad (1)$$

where $2\theta$ is the angle between beams 1 and 3. Here we consider the case of degenerate FWM, and thus all wave vectors were normalized to unity and indicate only the direction of propagation of the fields and not their wavelength. For a nondegenerate case, the extension is straightforward.   The delay between any two beams arriving at the point $\vec{r} = (x, y, z)$ is given by: $\tau_{i,j}(\vec{r}) = \vec{r} \cdot (\vec{k}_i - \vec{k}_j) c^{-1} + T_i - T_j$, where $T_i$ and $T_j$ are delays (provided externally) relative to an arbitrary zero which defines the time when all three pulses coincide in the "center" of the interaction region, and $c$ is the speed of light in the medium. From this, one can immediately deduce that the delays between the pulses can be mapped to the spatial coordinates as follows:

$$\tau_{3,1}(\vec{r}) = -2yc^{-1}\sin\theta + T_3 - T_1,$$
$$\tau_{2,1}(\vec{r}) = (z-y)c^{-1}\sin\theta + T_2 - T_1, \qquad (2)$$
$$\tau_{2,3}(\vec{r}) = (z+y)c^{-1}\sin\theta + T_2 - T_3.$$

If the fundamental beams are not in resonance with electronic transitions of the molecule, the signal will be observed only if two pulses (pump and Stokes) arrive simultaneously to establish coherence in the sample, to be probed later by the delayed third pulse. Due to phase matching considerations, the Stokes pulse should be provided by beam 2, while each of the two other pulses can serve either as pump or as probe. Therefore, we expect the picture to be symmetric with respect to the x-z plane at y = 0 (Figure 2).

Right of center, (the +y direction) pulse 3 precedes pulse 1, and thus should serve as a pump, and similarly, left of center (the –y direction) pulse 1 serves as a pump. According to equations (2) pulses 2 and 3 temporally coincide along the line $z = -y$, and pulses 2 and 1 arrive simultaneously to points along the line $z = y$. Thus, in the present arrangement, signals may be observed in two regions, as depicted by the time flow diagrams below the figure: right and below center, along the $z = -y$ line (pumped by pulse 3 and probed by pulse 1), and left and below center along the $z = y$ line (pumped by pulse 1 and probed by pulse 3). At nominally zero delay, the coherence peak appears at y=z=0 and the region above it (z>0) does not contain any signal, because in this region the probe precedes the excitation pair.

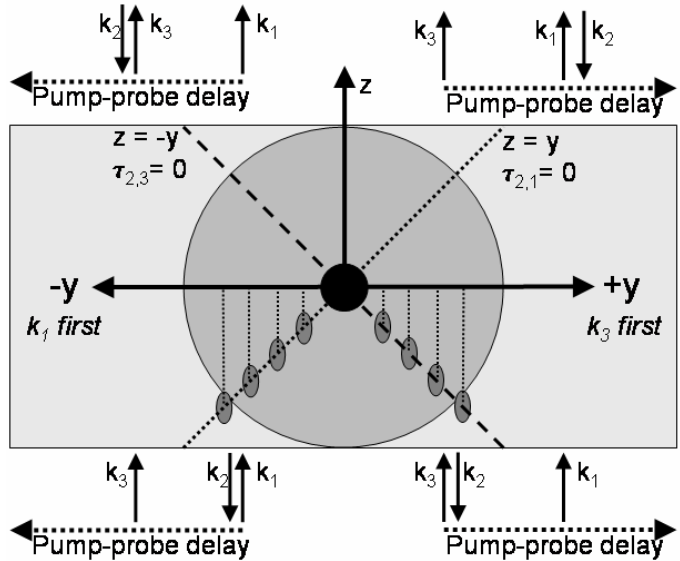

**Figure 2 schematic y-z slice of the interaction region. The outer rectangle represents the intersection between the $k_1$ and $k_3$ beams. The large gray circle represents beam $k_2$ The inner black circle depicts region where all three pulses coincide and DFWM signal is produced. Dashed and dotted lines represent coincidence between pulses 2 and 3 and pulses 2 and 1 respectively. Gray ellipses represent time delayed vibrational signal. The four pulse sequence schemes depict order of arrival of the pulses in each quadrant of the interaction volume.**

All points in the interaction region emit the FWM signal in the same, phase matched, direction, and the camera collects the entire spatially resolved CARS signal. The interpretation of the image requires calibration of the spatial coordinates in terms of angles and time delays between the pulses. A convenient method for calibrating the time–to-space mapping is to introduce external, known delays to the individual beams and observe the 'motion' (in the image plane) of the coherence peak as a function of these external delays. The coherence peak is significantly stronger than the time delayed signals and is very easy to observe due to various contributions to the nonlinear susceptibility. Once the image plane is calibrated, it is possible to move the coherence peak out of the picture by delaying the probe beam, thus enabling full utilization of the dynamic range of the camera. Moving the coherence peak breaks the symmetry between right and left, and in our experiments, the signal was collected along the calibrated $z = -y$ axis of the image (the sample is pumped by $k_3$, dumped by $k_2$, and probed by $k_1$).

A closely related configuration has been discussed by Naumov et al.[24, 25] who had theoretically analyzed configurations involving chirped pulses in a three dimensional beam crossing geometry, which allow for spatial mapping of different spectral components of the CARS signal. The proposed mapping transformation is based on the dispersion of the arrival time of different frequency components of linearly chirped pulses due to group velocity variations.

To demonstrate the method, we measured the spectra of the low-lying vibrational modes of bromoform ($CHBr_3$) and chloroform ($CHCl_3$). The measurements were performed on neat liquids in a 5 cm long quartz cell. The laser pulses were derived from a chirped pulse amplifier system providing pulses centered around 800 nm, ~70 fsec long, and of total energy up to 500 microJoul per pulse. In our three dimensional phase matching arrangement, the crossing angles were 2.3° and the beams' diameter at the crossing was ~5 mm.

To demonstrate the capabilities of the method, Figure 3a. depicts the image captured from a **single pulse** (no averaging). The dashed line is the calibrated $z = -y$ axis of the image (see above). The signal was measured along 50 parallel lines in the central region, covering the pulse duration, digitized and averaged for each value of the pump-probe delay. The inset of Figure 3b depicts the time domain signal and its power spectrum.

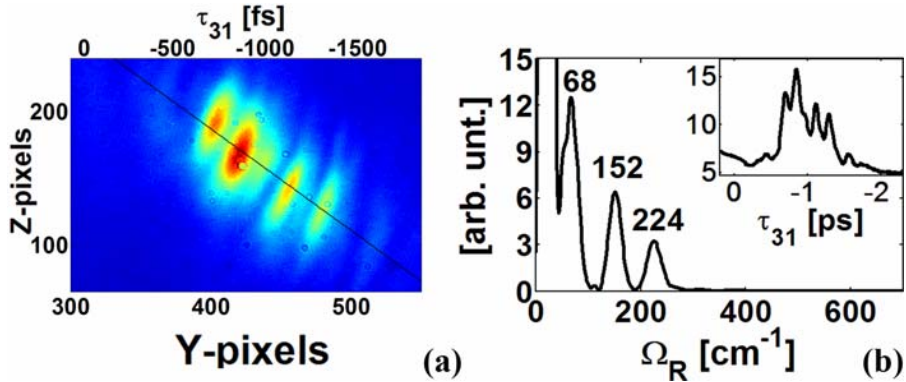

**Figure 3** Single-Pulse CARS image of bromoform "as captured". The line represents the calibrated $\tau_{2,3}=0$ line. The upper axis depicts the $\tau_{3,1}$ delay (pump-probe delay) (a). Power spectrum of the averaged time-domain signal. The time-domain signal is depicted in the inset (b).

The power spectrum shows the peaks at 152 cm$^{-1}$ and 224 cm$^{-1}$ corresponding to the asymmetric and symmetric bending of C-Br bonds respectively [26]. The third peak at 68 cm$^{-1}$ corresponds to the beat frequency between these two modes. The relatively weaker amplitude of the fundamental peaks stems from the limited bandwidth of the exciting pulses, and a detailed analysis of the terms contributing to the nonlinear susceptibility $\chi^{(3)}$, to be discussed in the future.

Figure 4 depicts the equivalent results for chloroform, where now the captured picture is an average over 2000 pulses. Figure 4**a** depicts the spatial image, and Figure 4**b** shows the time domain signal and the corresponding power spectrum. The observed peaks at 263 cm$^{-1}$ and 364 cm$^{-1}$ are assigned to asymmetric and symmetric bending of the C-Cl bonds[26]. The intensity of the 364 peak is weak due to the limited bandwidth of the fundamental pulses. The difference frequency homodyne peak is observed at 101 cm$^{-1}$.

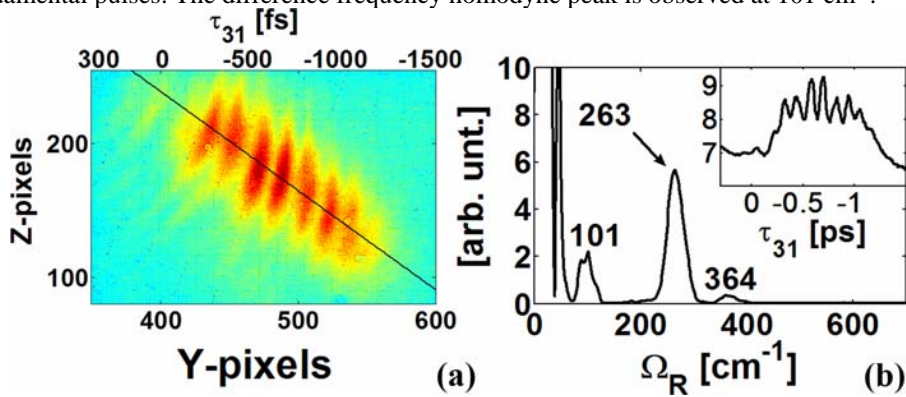

**Figure 4 Single-Shot CARS image of chloroform "as captured". The line represents calibrated $\tau_{2,3}=0$ line. The upper axis depicts the $\tau_{3,1}$ delay (pump-probe delay) (a). Power spectrum of the averaged temporal signal of single-shot CARS image of chloroform molecules. Averaged temporal signal is provided in insertion (b).**

In conclusion, we have demonstrated a new approach to single-shot time-resolved CARS. The method is simple to operate and enables the direct monitoring of several picoseconds of field-free vibrational evolution in one single image. The method was demonstrated on bromoform and chloroform, where the CARS spectrum of the low lying vibrational modes was measured. The ability to simultaneously measure both pump-Stokes and pump-probe delay times opens up the way to two- dimensional measurement of the coupling between ground and electronically excited states. Future development of this spectroscopic method involves the use of temporally and spectrally shaped pulses for selective excitation, as well as spatial filtering of the signal for selective detection of particular molecular degrees of freedom. Experiments are under way to explore these possibilities.

We gratefully acknowledge support form the James Franck program on light matter interaction and the Israel Science Foundation.